\documentclass[preprint,aps]{revtex4}

\usepackage{graphicx}

\begin{document}

\title{Dichotomy of Electronic Structure and Superconductivity between Single-Layer and Double-Layer FeSe/SrTiO$_3$ Films}
\author{Xu Liu$^{1,\sharp}$, Defa Liu$^{1,\sharp}$, Wenhao Zhang$^{3,4,\sharp}$, Junfeng He$^{1,\sharp}$, Lin Zhao$^{1}$, Shaolong He$^{1}$, Daixiang Mou$^{1}$,  Fansen Li$^{4}$, Chenjia Tang$^{3,4}$, Zhi Li$^{4}$, Lili Wang$^{4}$,  Yingying Peng$^{1}$, Yan Liu$^{1}$, Chaoyu Chen$^{1}$, Li Yu$^{1}$, Guodong  Liu$^{1}$,  Xiaoli Dong$^{1}$, Jun Zhang$^{1}$, Chuangtian Chen$^{5}$, Zuyan Xu$^{5}$,  Xi Chen$^{3}$, Xucun Ma$^{4,*}$,  Qikun Xue$^{3,*}$, and X. J. Zhou$^{1,2,*}$
}

\affiliation{
\\$^{1}$National Lab for Superconductivity, Beijing National Laboratory for Condensed Matter Physics, Institute of Physics,
Chinese Academy of Sciences, Beijing 100190, China
\\$^{2}$Collaborative Innovation Center of Quantum Matter, Beijing, China
\\$^{3}$State Key Lab of Low-Dimensional Quantum Physics, Department of Physics, Tsinghua University, Beijing
100084, China
\\$^{4}$Beijing National Laboratory for Condensed Matter Physics, Institute of Physics,
Chinese Academy of Sciences, Beijing 100190, China
\\$^{5}$Technical Institute of Physics and Chemistry, Chinese Academy of Sciences, Beijing 100190, China
}
\date{February 6, 2014}


\maketitle

{\bf
The latest discovery of possible high temperature superconductivity in the single-layer FeSe film grown on a SrTiO$_3$ substrate\cite{xueSTO}, together with the observation of its unique electronic structure and nodeless superconducting gap\cite{DFLiu,SLHe,DLFeng},  has generated much attention\cite{TXiang,DHLee,Timur,FZheng,NMBorisenko,JWang,DLFengN,theoryCao,CWChu,ZXShen}. Initial work also found that, while the single-layer FeSe/SrTiO$_3$ film exhibits a clear signature of superconductivity, the double-layer FeSe/SrTiO$_3$ film shows an insulating behavior\cite{xueSTO}. Such a dramatic difference  between the single-layer and double-layer FeSe/SrTiO$_3$ films is surprising and the underlying origin remains unclear. Here we report our comparative study between the single-layer and double-layer FeSe/SrTiO$_3$ films by performing a systematic angle-resolved photoemission study on the samples annealed in vacuum.  We find that, like the single-layer FeSe/SrTiO$_3$ film, the as-prepared double-layer FeSe/SrTiO$_3$ film is insulating and possibly magnetic, thus establishing a universal existence of the magnetic phase in the FeSe/SrTiO$_3$ films.  In particular, the double-layer FeSe/SrTiO$_3$ film shows a quite different doping behavior from the single-layer film in that it is hard to get doped and remains in the insulating state under an extensive annealing condition.  The difference originates from the much reduced doping efficiency in the bottom FeSe layer of the double-layer FeSe/SrTiO$_3$ film from the FeSe-SrTiO$_3$ interface. These observations provide key insights in understanding the origin of superconductivity and the doping mechanism in the FeSe/SrTiO$_3$ films. The property disparity  between the single-layer and double-layer FeSe/SrTiO$_3$ films may facilitate to fabricate electronic devices by making superconducting and insulating components on the same substrate under the same condition.
}

The FeSe superconductor is peculiar among the recently discovered iron-based superconductors\cite{JohnstonReview,FeSCReview,StewartReview} because it has the simplest crystal structure solely made of the FeSe layers that are believed to be one of the essential building blocks of the iron-based superconductors\cite{WuFeSe}. It also exhibits unique superconducting properties where the superconducting transition temperature (T$_c$) can be dramatically enhanced from  $\sim$8 K at ambient pressure to nearly 37 K under high pressure\cite{FeSePressure}. The latest indication of even higher T$_c$$\sim$80 K in the single-layer FeSe films grown on a SrTiO$_3$ substrate (Fig. 1a)\cite{xueSTO}, together with the observations of its unique electronic structure and nodeless superconducting gap\cite{xueSTO,DFLiu,SLHe,DLFeng},  has generated much interest\cite{TXiang,DHLee,Timur,FZheng,NMBorisenko,JWang,DLFengN,theoryCao,CWChu,ZXShen}. Surprisingly, it was found from the scanning tunneling microscopy(STM)/spectroscopy(STS) measurements that, on the given FeSe/SrTiO$_3$ film, while the single-layer film shows clear coherent peaks that are indicative of superconductivity, the double-layer film (Fig. 1b) shows a semiconducting/insulating behavior\cite{xueSTO}.  Such a disparate behavior between the single-layer and double-layer FeSe/SrTiO$_3$ films is rather puzzling. In a conventional picture, superconductivity is usually suppressed with the decrease of the film thickness due to dimensionality effect and may disappear when the thickness gets to a few layers or one layer, as demonstrated by the Pb films grown on a silicon substrate\cite{Xuelead} and the FeSe films grown on the graphene substrate\cite{xuegraphene}.  Even if one assumes that superconductivity arises only from the interface between the bottom FeSe layer and the SrTiO$_3$ substrate, it remains hard to understand why the top FeSe layer does not become superconducting because of the proximity effect. It is thus significant to understand the electronic origin on the obvious disparity between the single-layer and double-layer FeSe/SrTiO$_3$ films. Understanding this difference is important to uncover the origin of superconductivity in the single-layer FeSe/SrTiO$_3$ film  and may be helpful in further enhancement of superconducting T$_c$ in the FeSe/SrTiO$_3$ films.

In this paper, we report a comparative investigation between the single-layer and  double-layer FeSe/SrTiO$_3$ films by carrying out a systematic angle-resolved photoemission (ARPES) study on the samples annealed in vacuum.  First we find that, like the single-layer FeSe/SrTiO$_3$ film,  the as-prepared double-layer FeSe/SrTiO$_3$ film is insulating and possibly magnetic. These observations have established for the first time a universal existence of the magnetic state in the FeSe/SrTiO$_3$ films.  Second, we reveal that the double-layer FeSe/SrTiO$_3$ film shows a quite different doping behavior from the single-layer film. It is hard to get doped, and remains in the insulating state even after an extensive annealing process.  The difference originates from  the much reduced doping efficiency in the bottom FeSe layer of the double-layer FeSe/SrTiO$_3$ film from the FeSe-SrTiO$_3$ interface. Third, we have shown that, with sufficient annealing, the double-layer FeSe/SrTiO$_3$ film may follow similar doping trend as in the single-layer film\cite{SLHe}, although superconductivity has not been realized.  These observations provide  key insights in understanding the origin of superconductivity and the doping mechanism of the FeSe/SrTiO$_3$ films.  They also point to a possibility to further enhance superconductivity in the double-layer and multiple-layer FeSe/SrTiO$_3$ films. The dramatic difference between the single-layer and double-layer FeSe/SrTiO$_3$ films also provides a layer-engineering technique to fabricate electronic devices by making metallic, superconducting and insulating components on the same substrate under the same preparation conditions.

The single-layer and double-layer FeSe thin films were grown on a SrTiO$_3$ (001) substrate by the molecular beam epitaxy (MBE) method\cite{xueSTO}. To improve the sample quality, the growth procedure usually involves two steps: the first step to grow the film at a relatively low temperature, and the second step to anneal the as-prepared sample in vacuum at a relatively high temperature (see Supplementary and \cite{xueSTO} for details). It has been shown that, for the single-layer FeSe/SrTiO$_3$ film during the MBE growth,  the as-prepared sample grown at a relatively low temperature is not superconducting; the sample has to be post-annealed at a relatively high temperature to become superconducting\cite{xueSTO,SLHe,wenhao}. The detailed ARPES measurements keeping track on the electronic evolution of the single-layer FeSe/SrTiO$_3$ film from the initial non-superconducting sample to the superconducting sample revealed that,  the as-grown sample is mainly composed of a magnetic-like phase named ``N phase" while the superconductivity appears in the ``S phase"with sufficient electron doping\cite{SLHe,JFHeFeSe}. The N phase and S phase exhibit distinct electronic structures and they compete  and coexist in the intermittent annealing process\cite{SLHe}.

The as-prepared double-layer FeSe/SrTiO$_3$ film exhibits electronic structures that are similar to the as-prepared single-layer FeSe/SrTiO$_3$ film, indicating its possible magnetic state at low temperature.  Figure 1c shows a typical ``Fermi surface" of the as-prepared single-layer FeSe/SrTiO$_3$ film in the pure N phase (here we use ``Fermi surface" in a loose sense in that it corresponds to the maximum intensity contour in the spectral distribution in the momentum space). The corresponding band structure across the M3 cut and its temperature dependence are shown in Fig. 1d. Two hole-like bands are present at low temperatures: One is close to the Fermi level labeled as N2 band, and the other is nearly 110 meV below the Fermi level labeled as N3 band. With increasing temperature, the N2 band gets weaker and becomes invisible at 250 K.  In the meantime, the N3 band moves closer to the Fermi level with increasing temperature. The as-prepared double-layer FeSe/SrTiO$_3$ film shows rather similar electronic behavior as that of the as-prepared single-layer FeSe film. Its  ``Fermi surface"  (Fig. 1e) is also characterized by four ``strong spots" near M point and remnant weak spectral weight near $\Gamma$ point. Its band structure near M3 (Fig. 1f) exhibits two hole-like bands at low temperatures, with one N2 band close to the Fermi level and the other N3 band nearly 70 meV below the Fermi level at 18 K. The N2 band gets weaker with the increasing temperature and nearly disappears at 180 K.  The electronic structure of the as-prepared single-layer and double-layer FeSe/SrTiO$_3$ films shows a clear resemblance to that of the parent compound BaFe$_2$As$_2$ in its magnetic state (Fig. 1g and 1h)\cite{GDLiuBFA},  in terms of the ``Fermi surface" near the M point,  the band structure near the M point, and its temperature dependence.  They are also very similar to that of the multiple-layer FeSe/SrTiO$_3$ films\cite{DLFeng}.  All these observations indicate that the N phase in the as-prepared single-layer and double-layer FeSe/SrTiO$_3$ films is mostly likely magnetic, as indicated before\cite{SLHe,DLFeng}. The present observation of the N phase in the double-layer FeSe/SrTiO$_3$ film, together with our previous observation of the N phase in the single-layer FeSe/SrTiO$_3$ film\cite{SLHe}, has made a unified picture for its universal existence in the FeSe/SrTiO$_3$ films because the pure N phase was previously observed only in the triple and multiple-layer FeSe/SrTiO$_3$ films\cite{DLFeng}.  The magnetic transition temperature of the N phase in the single-layer and double-layer FeSe/SrTiO$_3$ films, determined from the temperature dependence of the band structure across M3 (Fig. 1d and 1f), is shown in Fig. 1i, together with the temperature for the multiple-layer FeSe/SrTiO$_3$ films\cite{DLFeng}.  It follows the similar trend found for the multiple-layer FeSe/SrTiO$_3$ films\cite{DLFeng}, i.e., it decreases with the increasing number of the FeSe layers. The single-layer and double-layer FeSe/SrTiO$_3$ films are peculiar in that their N phase exhibits the highest magnetic transition temperature among all the FeSe/SrTiO$_3$ films.

The distinct behaviors between the single-layer and double-layer FeSe/SrTiO$_3$ films show up when the as-prepared samples are post-annealed in vacuum under a similar condition. It was shown before that the as-prepared single-layer FeSe/SrTiO$_3$ film, after annealing in vacuum, can transform completely from an initial pure N phase into the other pure S phase\cite{SLHe}, with its  Fermi surface and band structure along typical momentum cuts shown in Fig. 2 (corresponding to the annealing Sequence 10 in \cite{SLHe}).  The Fermi surface of the S phase single-layer FeSe/SrTiO$_3$ film (Fig. 2a) is characterized by an electron-like Fermi surface around the M points while there is no indication of Fermi surface near the $\Gamma$ point\cite{DFLiu,SLHe}. Its corresponding band structure near $\Gamma$  point shows a hole-like band with its top nearly 80 meV below the Fermi level (Fig. 2b) and there is an electron-like band crossing the Fermi level near M (Fig. 2c and 2d)\cite{DFLiu}. For this particular annealed single-layer FeSe/SrTiO$_3$ film, it shows a superconducting gap that closes near 40 K\cite{SLHe}.  However, under an identical annealing condition, the double-layer FeSe/SrTiO$_3$ film remains in the N phase, with its Fermi surface (Fig. 2e) and band structure (Fig. 2f-h) similar to those of the as-prepared double-layer FeSe/SrTiO$_3$ film. In this case, the sample remains in the magnetic state and insulating. This result is consistent with the previous STM observation that, in the same FeSe/SrTiO$_3$ film, while the single-layer area shows a superconducting coherence peak with its position at 20.1 meV, the double-layer area shows an insulating behavior\cite{xueSTO}. Both the present ARPES results and previous STM results\cite{xueSTO} indicate that the double-layer FeSe/SrTiO$_3$ film indeed behaves quite differently from its single-layer counterpart if prepared and annealed under similar conditions.

One immediate question arising is whether one can transform the double-layer FeSe/SrTiO$_3$ film into the S phase, and even make it superconducting, by enhancing the annealing condition, as has been done for the single-layer FeSe/SrTiO$_3$ film\cite{SLHe}. The annealing condition can be enhanced in two ways: increasing the annealing time or elevating the annealing temperature. We used both ways to anneal the double-layer FeSe/SrTiO$_3$ film on two separate samples and the results are shown in Fig. 3 (for the sample \#1)  and Fig. 4 (for the sample \#2) (see Supplementary for the details of the annealing conditions).  Fig. 3a shows the ``Fermi surface" evolution of the double-layer FeSe/SrTiO$_3$ film (sample \#1) with the increase of annealing time at a constant annealing temperature of 350 $^{\circ}$C. The corresponding evolution of the band structure along three typical cuts is shown in Fig. 3(b-d). With the increasing annealing time, the initial N phase bands get weaker but are still present until the last annealing sequence 6 which already has a long annealing time over 90 hours. New features corresponding to the S phase (S1, S2 and S3 bands marked in Fig. 3) start to emerge in the sequence 5 and become clear for the sequence 6. During the annealing process, the double-layer FeSe/SrTiO$_3$ film undergoes a transition from the N phase into the S phase. This trend of evolution is rather similar to that observed in the single-layer FeSe/SrTiO$_3$ sample but it corresponds to an early annealing stage where the major N phase coexists with the emergent minor S phase\cite{SLHe}. Fig. 3e shows the band structure across M2 at a low (left panel) and a high (right panel) temperatures for the sample \#1 for the last sequence 6. The Fermi distribution function is removed in order to highlight a possible gap opening near the Fermi level.  The symmetrized EDCs near the underlying Fermi momentum measured at different temperatures are shown in Fig. 3f. An energy gap of $\sim$18 meV is observed at a low temperature of 23 K, which manifest itself as the appearance of a dip in the symmetrized EDCs (Fig. 3f) and a suppression of the spectral weight near the Fermi level in the photoemission image (Fig. 3e). The EDC peak is broad. The gap size shows little change with temperature and does not close until 70 K (Fig. 3f and Fig. 3e). All these observations indicate that the observed energy gap is an insulating gap.  This is consistent with the case in the single-layer FeSe/SrTiO$_3$ film where it is found that, in the early  annealing stage when the S phase just appears and has a low doping level, it opens an insulating gap\cite{JFHeFeSe}.

Figure 4 shows the Fermi surface and band structure evolution of the double-layer FeSe/SrTiO$_3$ film (sample \#2) under another annealing condition to gradually increase the annealing temperature.  To avoid redundancy, we started with the initial sample that is already annealed for a couple of cycles at lower temperatures and then annealed at 350 $^{\circ}$C (sequence 1 for sample \#2). It shows dominant N phase features, and also some signatures of the S phase. With further annealing, the bands corresponding to the N phase (N1, N2 and N3 bands) get weaker while the S phase bands (S1, S2 and S3 bands) get stronger. When it comes to the sequences 5 and 6, the N phase disappears and the measured electronic structure is dominated by the pure S phase. Such a Fermi surface and band structure evolution follows a rather similar trend as that for the single-layer FeSe/SrTiO$_3$ sample\cite{SLHe}.  It appears that higher annealing temperature is more effective in giving rise to the transition from the N phase to the S phase even though the annealing time is much shorter than that used for the sample \#1 (Fig. 3). Fig. 4e shows the band structure measured at different temperatures along the M2 cut for the sequence 6. The corresponding symmetrized EDCs measured at different temperatures on the Fermi momentum are shown in Fig. 4f. One can see a sharp coherence peak,  and a clear gap opening at low temperature with a size of 13.5 meV which gets closed near 55 K. These observations are similar to that observed in the single-layer FeSe/SrTiO$_3$ film\cite{DFLiu,SLHe,DLFeng} which indicate that the sample for the sequence 6 is superconducting.

During the annealing of the double-layer FeSe/SrTiO$_3$ films, we realized that the situation can be more complicated than annealing the single-layer FeSe/SrTiO$_3$ films. In addition to the doping effect that is involved, there is also a possibility of FeSe evaporation during the annealing process. For the single-layer FeSe/SrTiO$_3$ film, we can always probe the electronic structure of the single-layer FeSe film because the bare SrTiO$_3$ substrate does not contribute obvious signal\cite{SLHe,DLFeng} even if some portion of the FeSe layer is evaporated. However, in the case of the double-layer FeSe/SrTiO$_3$ film, there is a chance that the top layer may evaporate and the sample may gradually become mixed with double-layer and single-layer regions.   To examine on this possibility, we checked the FeSe/SrTiO$_3$  films by using STM before and after annealing under different conditions.   As shown in Fig. S3 in the Supplementary, when the annealing temperature is relatively low (450 $^{\circ}$C in this case), after annealing for 15 hours, the double-layer region remains nearly intact with a similar occupation ratio before and after annealing.  However, when the annealing temperature is increased to 500 $^{\circ}$C, after annealing for only 3 hours,  the area of the initial triple-layer region is obviously reduced, and single-layer region emerges that is absent in the original sample. This indicates that at a high annealing temperature ($\sim$500 $^{\circ}$C and above), significant FeSe evaporation can occur during the annealing process. Based on these STM results, we intentionally chose the annealing temperature of 350 $^{\circ}$C for the sample \#1 of the double-layer FeSe/SrTiO$_3$ film (Fig. 3). Since it is well below 450 $^{\circ}$C used in the STM experiment (Fig. S3a and S3b in Supplementary), the FeSe evaporation can be minimized so that it keeps its double-layer structure during the annealing process. One may argue whether some areas may still become single-layer due to the long annealing time in spite of a low annealing temperature, and whether the signal of the observed S phase may come from the exposed single-layer area, instead of an intrinsic property of the double-layer area. We believe this is unlikely because, if we have single-layer areas present, from our experience of annealing the single-layer FeSe/SrTiO$_3$ samples\cite{SLHe,JFHeFeSe}, these  single-layer areas should already become superconducting under the annealing conditions we used. This is not consistent with our observation of the broad photoemission spectra and an insulating gap in the S phase (Fig. 3f).   On the other hand,  although we observed a complete transformation of the N phase into the S phase and even the appearance of superconductivity in the sample \#2  of the double-layer FeSe/SrTiO$_3$ film (Fig. 4), from the STM results (Fig. S3 in Supplementary), it is very likely that this is due to the evaporation of the top FeSe layer that results in the formation of the single-layer FeSe regions during the annealing process. In other words, these observations reflect the behaviors of the single-layer FeSe/SrTiO$_3$ film, instead of the double-layer FeSe/SrTiO$_3$ film that we intended to probe.

The above results clearly indicate that the double-layer FeSe/SrTiO$_3$ film behaves quite differently from the single-layer case in that it is much harder to get doped.  Under the same annealing condition when the N phase in the as-prepared single-layer FeSe/SrTiO$_3$ film is completely transformed into a pure S phase and becomes superconducting, the  double-layer film remains mainly in the N phase (Fig. 2). Even after an enhanced annealing by extending the annealing time (Fig. 3), it only reached an early stage of the transformation where the S phase starts to emerge and has a low doping. The complete transformation from the N phase to the S phase requires to further enhance the annealing condition by elevating the annealing temperature as carried out on the sample \#2. However, it is complicated by the possible FeSe evaporation so that it is hard to probe the intrinsic doping evolution of the double-layer FeSe/SrTiO$_3$ film.

The present results of annealing the double-layer FeSe/SrTiO$_3$ film provide an important insight on the origin of superconductivity in the FeSe/SrTiO$_3$ films. It has been a prominent issue on where the superconductivity in the single-layer FeSe/SrTiO$_3$ film originates: from the FeSe layer alone, or the SrTiO$_3$ substrate, or the interface between the SrTiO$_3$ substrate and the FeSe layer.  The dichotomy between the single-layer and double-layer FeSe/SrTiO$_3$ films makes it clear that the superconductivity cannot originate from the FeSe layer alone. If the superconductivity comes only from the FeSe layer, one would expect that the double-layer FeSe/SrTiO$_3$ film, especially the top FeSe layer of the double-layer film,  would behave similarly like the single-layer FeSe/SrTiO$_3$ film to become superconducting under the same annealing condition. This result, together with the observation of the dramatically different behaviors between the FeSe films grown on the graphene substrate\cite{xuegraphene} and on the SrTiO$_3$ substrate\cite{xueSTO}, point to the significant role of the interface in giving rise to the superconductivity in the FeSe/SrTiO$_3$ films.

The present work on the double-layer FeSe/SrTiO$_3$ film, as well as that on the single-layer FeSe/SrTiO$_3$ film\cite{SLHe,JFHeFeSe}, clearly indicates the essential role of the carrier doping on the superconductivity in the FeSe/SrTiO$_3$ films: superconductivity occurs only when an appropriate amount of doping is introduced at the interface. The dichotomy of annealing the single-layer and double-layer FeSe/SrTiO$_3$ films provides further information on the doping mechanism of the FeSe/SrTiO$_3$ films. First, our observation of the pure N phase in the as-prepared single-layer and double-layer FeSe/SrTiO$_3$ films indicate that, even when the SrTiO$_3$ substrate is metallic after the heat-treatment at a high temperature accompanied by Se flushing\cite{xueSTO}, the FeSe/SrTiO$_3$ films are not automatically doped in the preparation stage at a relatively low temperature.  It is clear that the carrier doping is realized at a late stage when the sample is vacuum annealed at a relatively higher tempetaure\cite{xueSTO,SLHe}.  The pure N phase for the single-layer and double-layer FeSe/SrTiO$_3$ films was not observed before\cite{DLFeng} probably because their preparation temperature is higher than that used in our samples.  Second, the present study provides a clear clue on the origin of the carrier doping.  As proposed before, there are two possibilities of carrier doping in the vacuum annealing process of the FeSe/SrTiO$_3$ films. One is the evaporation of Se or formation of the Se vacancies in the FeSe films while the other is the formation of oxygen vacancies near the SrTiO$_3$ surface\cite{DFLiu,SLHe,DLFeng}.  Recent STM measurements on a vacuum annealed single-layer FeSe/SrTiO$_3$ film found that\cite{wenhao}, in the as-prepared FeSe film, there are some extra Se adatoms on the surface and the sample is insulating.  In the initial stage of annealing, these extra Se adatoms are gradually removed, and further annealing leads to the formation of Se vacancies. The annealing process gives rise to an insulator-superconductor transition that is in agreement with our ARPES measurements\cite{JFHeFeSe}.  If the doping process involves only the FeSe film itself, one would expect similar behavior between the single-layer and double-layer FeSe/SrTiO$_3$ films under the same annealing conditions, particularly the top layer of the double-layer FeSe/SrTiO$_3$ film experiences a similar environment as the single-layer FeSe/SrTiO$_3$ film so that one would expect they have similar doping effect.  This is apparently not consistent with the dramatically different annealing behaviors observed between the single-layer and double-layer FeSe/SrTiO$_3$ films (Fig. 2).   The present results make it clear that the doping process does not involve the FeSe films alone, i.e., the evaporation of Se during annealing is not a dominant factor in carrier doping. Instead, the SrTiO$_3$ substrate and the interface between the FeSe and SrTiO$_3$ must play a major role.

The dichotomy between the single-layer and double-layer FeSe/SrTiO$_3$ films can be naturally understood  when the carrier doping is realized by the charge transfer from the SrTiO$_3$ surface.  It has been shown that a two-dimensional electron gas can be formed on the SrTiO$_3$(001) surface even when it is annealed in vacuum at a rather low temperature ($\sim$300$^{\circ}$C)\cite{STO2D}.  As shown before, only when the S phase is formed and the doping level of the S phase exceeds a critical value ($\sim$0.089 electrons/Fe) can it become superconducting\cite{JFHeFeSe}.  In the single-layer FeSe/SrTiO$_3$ film, all the transferred charge from the SrTiO$_3$ surface can be taken by the sole FeSe layer and superconductivity is observed in the doping range over 0.089 electrons/Fe\cite{SLHe,JFHeFeSe}.  However, when it comes to the double-layer or multiple-layer FeSe/SrTiO$_3$ films, the same charge carriers from the SrTiO$_3$ surface will be shared by all the FeSe layers, rendering the doping of the bottom FeSe layer much reduced.   Here one question arises as to how the charge carries will be distributed among multiple FeSe layers and how much is left in the bottom layer.  The extreme case one may expect is that all the carriers remain confined only near the interface between the SrTiO$_3$ surface and the bottom FeSe layer.  If this is the case, in Figs. 2 and 3, one would expect that the bottom layer of the double-layer FeSe/SrTiO$_3$ film is completely transformed into the S phase and become superconducting, like in the single-layer FeSe/SrTiO$_3$ film.  The top FeSe layer may also become superconducting due to the proximity effect of the superconducting bottom layer.  With the photoemission probe depth on the order of 5$\sim$10 $\AA$ at the photon energy we used (21.2 eV), it is possible to see both layers that gives a mixed signal from the superconducting bottom layer and the undoped top layer.  This is apparently not consistent with the measured results in Fig. 2 and Fig. 3 where the signal is dominated by the N phase. Even though we observed some signal from the S phase after annealing,  it shows an insulating behavior instead of a superconducting behavior as expected.  It is therefore more plausible that the carriers from the charge reservoir of the SrTiO$_3$ surface is spread over all FeSe layers, but in an uneven manner:  the FeSe layer closer to the interface gets more doped while the FeSe layer farther away from the interface gets less doped.  This can explain why the double-layer or multiple-layer FeSe films are harder to get doped to become superconducting because the spread of the charge carriers over all the FeSe layers makes the doping on the bottom FeSe layer low that is not sufficient to reach the critical doping level as required in realizing superconductivity in the single-layer FeSe/SrTiO$_3$ film\cite{SLHe,JFHeFeSe}.  One may further expect that the thicker the FeSe/SrTiO$_3$ film is, the harder it is to dope, and the easier it is to stay as an N phase. Because ARPES and STM are surface-sensitive techniques that probe mainly the top one or two layers which are least doped, this explains why one sees robust magnetic ground state in the double-layer film here and in the multiple-layer FeSe/SrTiO$_3$ films\cite{DLFeng}.

The present work provides clues on understanding a number of puzzling issues associated with the superconductivity in the FeSe/SrTiO$_3$ films and the bulk FeSe samples. First, it sheds light on the important issue about the  nature of the parent compound for the bulk FeSe superconductor.  It is clear that the MBE-grown FeSe/SrTiO$_3$ films behave differently from the bulk FeSe in that the FeSe/SrTiO$_3$ films exhibit a universal magnetic state (N phase) that is absent in the bulk FeSe sample. The difference can be attributed to the delicate difference in the sample composition. In the MBE-grown FeSe/SrTiO$_3$ films, the atomic control of the composition makes it possible to prepare FeSe$_{1+x}$  or FeSe$_{1-x}$ films that are nearly perfect for the Fe ordering although there may be Se adatoms or Se vacancies formation\cite{wenhao}. On the other hand, for the bulk Fe$_{1+x}$Se superconductor, slight excess Fe is inevitable and necessary for stabilizing the crystal structure\cite{WuFeSe,WBao} while iron vacancies can be formed in the nano-form of the Fe$_{1-x}$Se compounds\cite{TKChen}. The discovery of the magnetic N phase in the as-prepared FeSe/SrTiO$_3$ films, and its transformation into the S phase and becoming superconducting via doping, make it clear that this magnetic N phase,  instead of FeTe commonly believed\cite{FeTe}, can be the parent compound for the family of the FeSe superconductor.
These results also naturally address another puzzle why the multiple-layer FeSe/SrTiO$_3$ films stay at the magnetic state, even with more than 50 FeSe layers\cite{DLFeng},  but do not recover to the bulk superconducting state with a T$_c$ around 8 K\cite{WuFeSe}. In this case, in addition to a delicate composition difference between the  MBE-grown FeSe/SrTiO$_3$ films and the bulk FeSe superconductor, for the MBE-grown multiple-layer FeSe/SrTiO$_3$ films, the carrier doping transferred from the SrTiO$_3$ surface is not high enough to induce bulk superconductivity.

The present study points to a new opportunity in enhancing superconductivity in the double-layer or multiple-layer FeSe/SrTiO$_3$ films. Our present study has clearly shown that the electronic evolution of the double-layer FeSe/SrTiO$_3$ film shows a tendency to follow a similar route as in the single-layer film (emergence of the S phase with annealing, Fig. 3). It is now clear that the difficulty in achieving a metallic or superconducting state in the double-layer and multiple-layer FeSe/SrTiO$_3$ films lies in how to effectively dope the FeSe layers into an appropriate doping level. As it is common in the high temperature cuprate superconductors that the existence of multiple-CuO$_2$ layers in one unit cell may exhibit higher superconducting transition temperature (T$_c$) than its single-layer counterparts\cite{EisakiPRB}, it is interesting to explore whether the double-layer FeSe/SrTiO$_3$ film, or the multiple-layer FeSe/SrTiO$_3$ films, may become superconducting and possess even higher T$_c$ than the single-layer film if doped properly.   To enhance the doping in the double-layer and multiple-layer FeSe/SrTiO$_3$ films, and to overcome the complication of the FeSe evaporation, one has to take a different approach from the present simple annealing process. This can be realized by either substitution within the FeSe films or chemical deposition on the top of the FeSe surface.  This provides an additional handle to tune superconductivity in the FeSe/SrTiO$_3$ films. It is expected that future study along this line will shed more insight on the origin of high temperature superconductivity in the FeSe/SrTiO$_3$ films.

In summary, by taking a systematic comparative ARPES study, we have uncovered distinct behaviors of the electronic structure and superconductivity between the single-layer and double-layer FeSe/SrTiO$_3$ films. We find that, like its single-layer FeSe/SrTiO$_3$ counterpart, the double-layer FeSe/SrTiO$_3$ film follows a similar tendency to transform from an initial N phase into an S phase when annealed in vacuum. But it is much harder for the double-layer FeSe/SrTiO$_3$ film to get doped in transforming completely from the N phase into the S phase and becoming superconducting. This difference lies in the much reduced doping efficiency in the bottom FeSe layer for the double-layer and multiple-layer FeSe/SrTiO$_3$ films. Our observations provide key insights in understanding the dichotomy between the single-layer and double-layer FeSe/SrTiO$_3$ films, and the difference between the MBE-prepared FeSe/SrTiO$_3$ films and the bulk FeSe superconductor. It also points to a  possibility in further enhancing superconductivity in the double-layer and multiple-layer FeSe/SrTiO$_3$ films. The realization of metallic, superconducting and insulating films on the same substrate and under the same preparation conditions also provides an opportunity of simple layer-engineering in making electronic devices.

$^{\sharp}$These people contribute equally to the present work.

$^{*}$Corresponding authors: XJZhou@aphy.iphy.ac.cn, qkxue@mail.tsinghua.edu.cn, xcma@aphy.iphy.ac.cn

\vspace{3mm}

\noindent {\bf Acknowledgement} XJZ thanks financial support from the NSFC (11190022,11334010 and 11374335) and the MOST of China (973 program No: 2011CB921703 and 2011CBA00110). QKX and XCM thank support from the MOST of China (program No. 2009CB929400 and  No. 2012CB921702).

\vspace{3mm}




\newpage

\begin{figure*}[tbp]
\begin{center}
\includegraphics[width=1.0\columnwidth,angle=0]{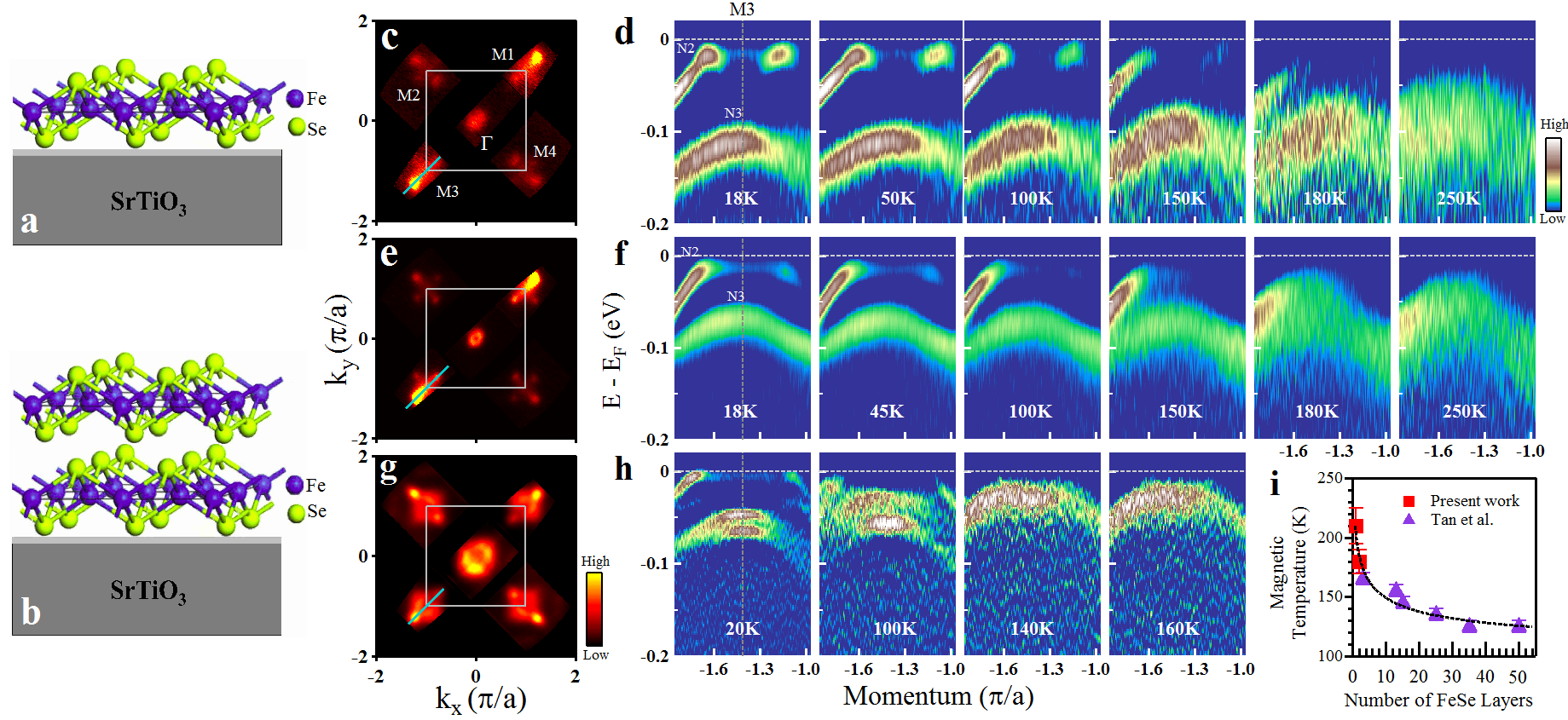}
\end{center}
\caption{Electronic structure and its temperature dependence of the as-prepared single-layer and double-layer FeSe/SrTiO$_3$ films and their comparison with that of BaFe$_2$As$_2$. (a) and (b) show  schematic structure of single-layer and double-layer FeSe films grown on SrTiO$_3$ (001) substrate, respectively.   (c). ``Fermi surface" of the as-prepared single-layer FeSe/SrTiO$_3$ film measured at 18 K obtained by integrating the spectral weight over a small energy window [-0.03eV,-0.01eV]. For convenience, we label the equivalent four M points as M1, M2, M3 and M4.  (d). Temperature dependence of the band structure measured along the M3 cut as shown in (c). The band structures shown here , as well as those in (f) and (h),  are the second derivative of the original bands with respect to the energy, in order to highlight the bands better\cite{SLHe}. One can see that the hole-like band which is obvious at low temperatures get weaker with increasing temperature and becomes invisible at 250 K. (e). ``Fermi surface" of the as-prepared double-layer FeSe films measured at 18 K. (d). Temperature dependence of the band structure measured along the M3 cut as shown in (e). The hole-like band disappears at $\sim$180 K. (g). Fermi surface of BaFe$_2$As$_2$ measured at 40 K which is below the magnetic transition temperature at $\sim$138 K\cite{GDLiuBFA}. (h). Temperature dependence of the band structure across the M3 point. (i). The transition temperature as a function of the number of FeSe layers in the FeSe/SrTiO$_3$ films. Here the transition temperature for the single-layer and double-layer FeSe films (red squares) is determined from the temperature dependence of the band structure ((d) and (f)) where the hole-like bands begin to disappear. The temperature for other films (violet triangles) with more FeSe layers is taken from Tan et al.\cite{DLFeng}.}
\end{figure*}

\begin{figure*}[tbp]
\begin{center}
\includegraphics[width=1.0\columnwidth,angle=0]{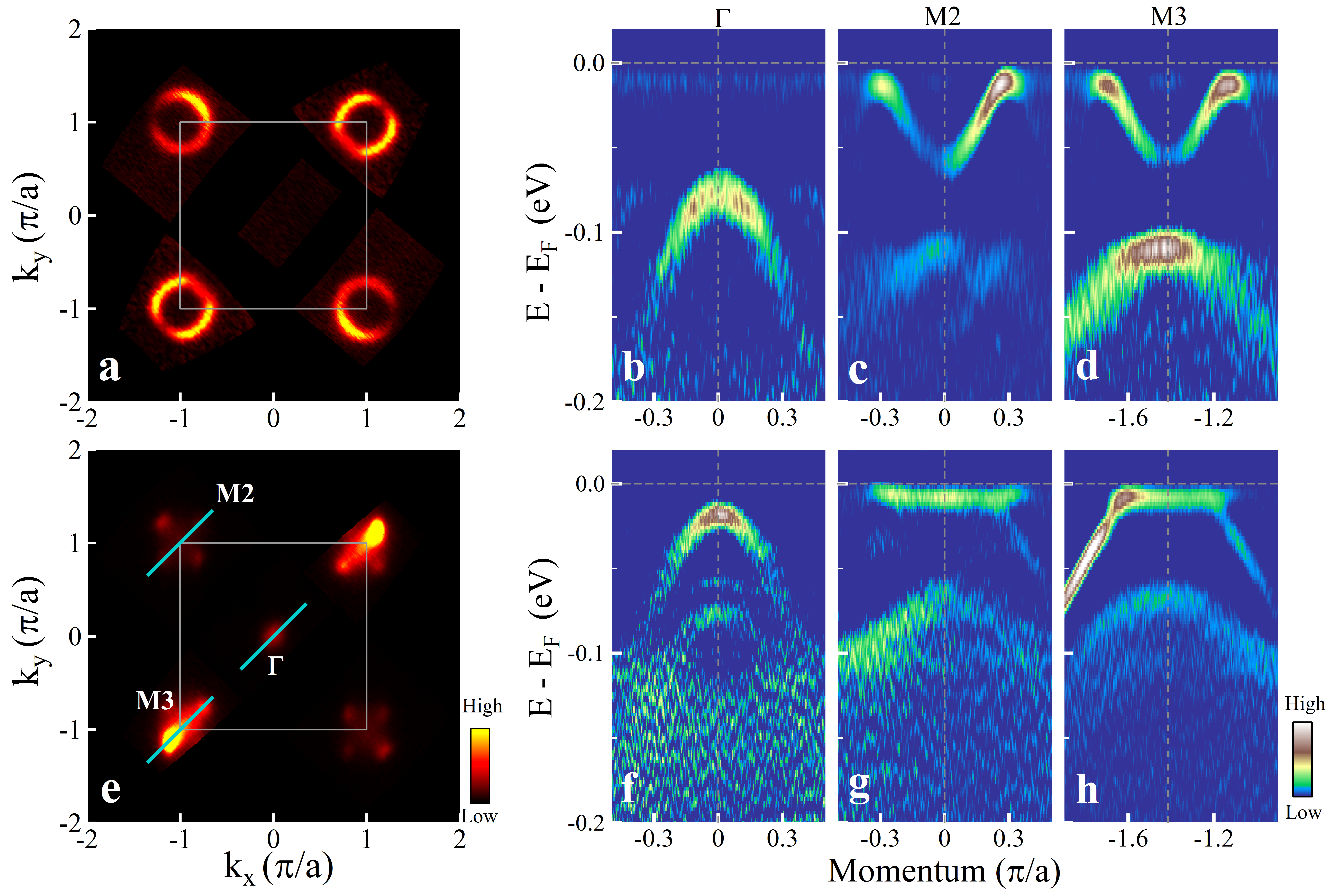}
\end{center}
\caption{Dichotomy of Fermi surface and band structure between the single-layer and double-layer FeSe/SrTiO$_3$ films after being annealed under the same condition. The annealing condition corresponds to the Sequence 10 in Ref.\cite{SLHe}. (a). Fermi surface of the annealed single-layer FeSe/SrTiO$_3$ film. (b-d). Corresponding band structure of the annealed single-layer FeSe/SrTiO$_3$ film along the $\Gamma$, M2 and M3 cuts, respectively. (e). Fermi surface of the annealed double-layer FeSe/SrTiO$_3$ film. (f-h). Corresponding band structure of the annealed double-layer FeSe/SrTiO$_3$ film along the $\Gamma$, M2 and M3 cuts, respectively. The band structures shown in (b,c,d) and (f,g,h) are second derivative of the original band with respect to energy.}
\end{figure*}

\begin{figure*}[tbp]
\begin{center}
\includegraphics[width=1.0\columnwidth,angle=0]{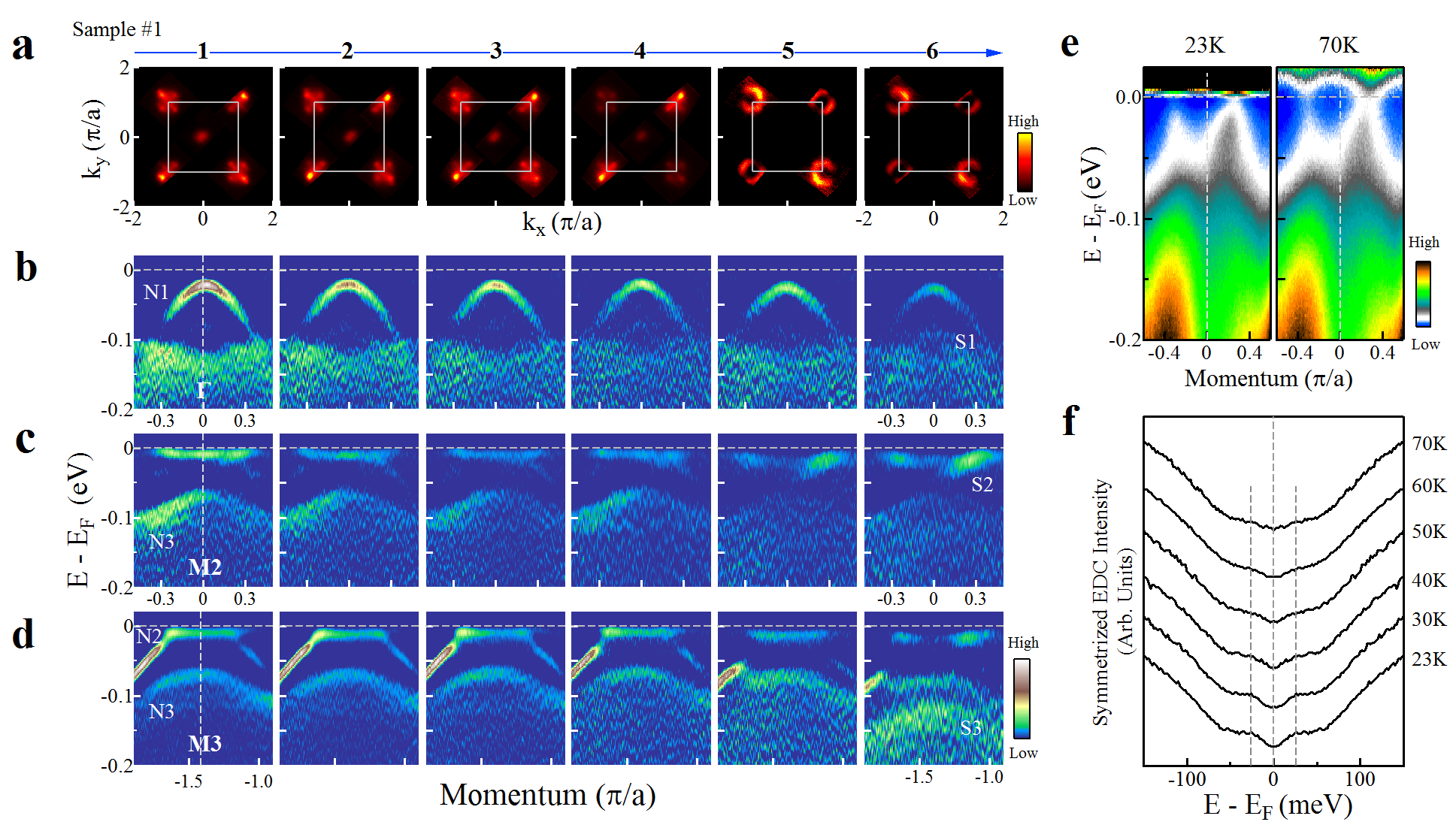}
\end{center}
\caption{Fermi surface and band structure evolution of the double-layer FeSe/SrTiO$_3$ film (sample $\#$1) annealed at a constant annealing temperature of 350$^{\circ}$C in vacuum for different times. The sequence 1 to 6 correspond to an accumulative time of 15,  30.5, 46.5,  66.5,  87.5, and 92.5 hours, respectively (see Supplementary for experimental details). (a). Fermi surface evolution as a function of the annealing time. (b-d) Band structure evolution with annealing time for the momentum cuts across $\Gamma$ (b), M2 (c) and M3 (d). The location of the three cuts are the same as shown in Fig. 2e. The band structures shown here are second derivative of the original images with respect to energy. (e). Band structure (original data) of the annealed double-layer FeSe/SrTiO$_3$ film (corresponding to the sequence 6) measured at 23 K (left panel) and 70 K (right panel). The photoemission images are divided by the corresponding Fermi distribution function to highlight opening or closing of an energy gap. (f). Corresponding symmetrized EDCs on the Fermi momentum measured at different temperatures.}
\end{figure*}

\begin{figure*}[tbp]
\includegraphics[width=1.0\columnwidth,angle=0]{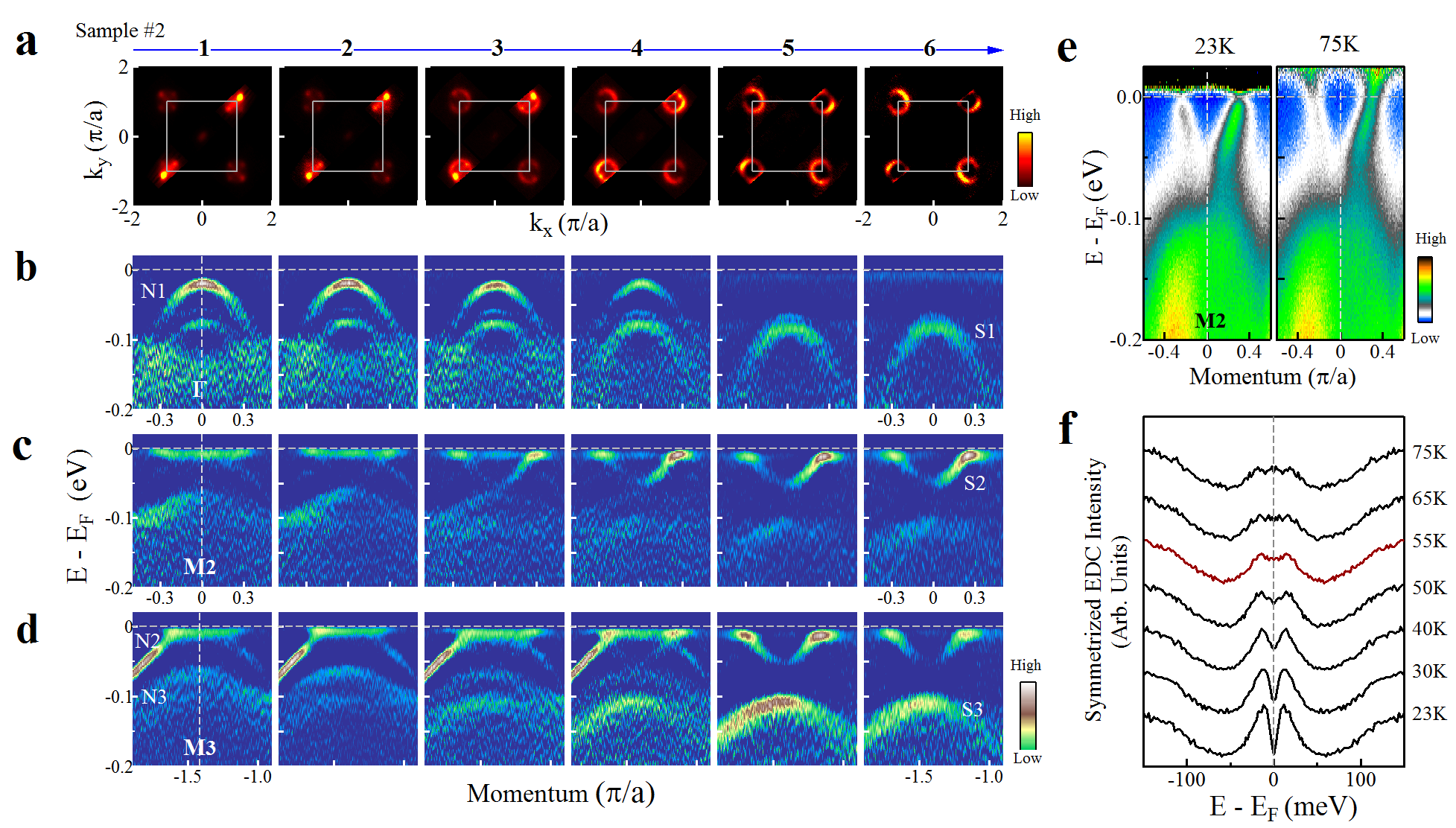}
\begin{center}
\caption{Fermi surface and band structure evolution of the double-layer FeSe/SrTiO$_3$ film (sample $\#$2) annealed with increasing annealing temperatures in vacuum. The highest annealing temperature is 350$^{\circ}$C for the sequence 1, 450$^{\circ}$C for the sequence 2, 500$^{\circ}$C for the sequence 3, 535$^{\circ}$C for the sequence 4, and then kept at 535$^{\circ}$C to increase the annealing time for the sequences 5 and 6. Refer to the Supplementary for further experimental details of the annealing process.  (a). Fermi surface evolution during the annealing process. (b-d) Band structure evolution for the momentum cuts across $\Gamma$ (b), M2 (c) and M3 (d). The location of the three cuts are the same as shown in Fig. 2e. The band structures shown here are second derivative of the original images with respect to energy. (e). Band structure (original data) of the annealed double-layer FeSe/SrTiO$_3$ film (corresponding to the sequence 6) measured at 23 K (left panel) and 75 K (right panel). The photoemission images are divided by the corresponding Fermi distribution function to highlight opening or closing of an energy gap. (f). Corresponding symmetrized EDCs on the Fermi momentum measured at different temperatures.}
\end{center}
\end{figure*}


\begin {thebibliography} {99}


\bibitem{xueSTO} Q. Y. Wang et al., Interface-induced high-temperature superconductivity in single unit-cell FeSe films on SrTiO$_3$. Chin. Phys. Lett. {\bf 29}, 037402(2012).
\bibitem{DFLiu} D. F. Liu et al., Electronic origin of high-temperature superconductivity in single-layer FeSe superconductor. Nat. Commun. {\bf 3}, 931(2012).
\bibitem{SLHe} S. L. He et al., Phase diagram and electronic indication of high-temperature superconductivity at 65 K in single-layer FeSe films. Nature Mater. {\bf 12}, 605 (2013).
\bibitem{DLFeng} S. Y. Tan et al., Interface-induced superconductivity and strain-dependent spin density wave in FeSe/SrTiO$_3$ thin films. Nature Mater. {\bf 12}, 6340 (2013).

\bibitem{TXiang}K. Liu et al., Atomic and electronic structures of FeSe monolayer and bilayer thin films on SrTiO$_3$(001): First-principles study. Phys. Rev. B {\bf 85}, 235123 (2012).
\bibitem{DHLee} Y. Y. Xiang et al., High-temperature superconductivity at the FeSe/SrTiO$_3$ interface. Phys. Rev. B {\bf 86}, 134508 (2012).
\bibitem{Timur} T. Bazhirov and M. Cohen, Effects of charge doping and constrained magnetization on the electronic structure of an FeSe monolayer. J. Phys.: Condens. Matter {\bf 25}, 105506 (2013).
\bibitem{FZheng} F. W. Zheng et al., Antiferromagnetic FeSe monolayer on SiTiO$_3$: The charge doping and electric field effects. Sci. Reports {\bf 3}, 2213 (2013).
\bibitem{NMBorisenko} S. Borisenko, Fewer atoms, more information. Nature Mater. {\bf 12}, 600 (2013).
\bibitem{JWang} W. H. Zhang et al., Direct observation of high temperature superconductivity in one-unit-cell FeSe films. Chin. Phys. Lett. {\bf 31}, 017401 (2014).
\bibitem{DLFengN} R. Peng et al., Enhanced superconductivity and evidence for novel pairing in single-layer FeSe on SrTiO$_3$ thin film under large tensile strain. arXiv:1310.3060 (2013).
\bibitem{theoryCao} H. Y. Cao et al., The interfacial effects on the spin density wave in FeSe/SrTiO$_3$ thin film. arXiv:1310.4024 (2013).
\bibitem{CWChu} L. Z. Deng et al., The Meissner and mesoscopic superconducting states in 1-4 unit-cell FeSe-films up to 80 K. arXiv:1311.6459 (2013).
\bibitem{ZXShen} J. J. Lee et al., Evidence for pairing enhancement in single unit cell FeSe on SrTiO$_3$ due to cross-interfacial electron-phonon coupling. arXiv:1312.2633 (2013).

\bibitem{JohnstonReview} D. C. Johnston, The puzzle of high temperature superconductivity in layered iron pnictides and chalcogenides. Advances in Physics {\bf 59}, 803 (2010).
\bibitem{FeSCReview}J. Paglione and R. L. Greene, High temperature superconductivity in iron-based superconductors, Nature Phys. {\bf 6}, 645 (2010).
\bibitem{StewartReview} G. R. Stewart, Superconductivity in iron compounds. Rev. Modern Phys. {\bf 83}, 1589 (2011).

\bibitem{WuFeSe} F. C. Hsu et al., Superconductivity in the PbO-type structure $\alpha$-FeSe. Proc. Natl. Acad. Sci. USA {\bf 105}, 14262 (2008).
\bibitem{FeSePressure} S. Medvedev et al., Electronic and magnetic phase diagram of $\beta$-Fe$_{1.01}$Se with superconductivity at 36.7 K
under pressure. Nature Mater. {\bf 8}, 630 (2009).

\bibitem{Xuelead} Y. Guo et al., Superconductivity modulated by quantum size effects. Science {\bf 306}, 1915 (2004).

\bibitem{xuegraphene} C. L. Song et al., Molecular-beam epitaxy and robust superconductivity of stoichiometric FeSe crystalline films on bilayer graphene. Phys. Rev. B {\bf 84}, 020503 (2011).

\bibitem{wenhao} W. H. Zhang et al., Interface charge doping effect on superconductivity of single unit-cell FeSe films on SrTiO$_3$ substrates, unpublished.

\bibitem{JFHeFeSe} J. F. He et al., Electronic evidence of an insulator-superconductor transition in single-layer FeSe/SrTiO$_3$ films, arXiv£º1401.7115 (2014).

\bibitem{GDLiuBFA} G. D. Liu et al., Band-structure reorganization across the magnetic transition in BaFe$_2$As$_2$ seen via high-resolution angle-resolved photoemission. Phys. Rev. B {\bf 80}, 134519 (2009).




\bibitem{STO2D} R. Di Capua et al., Observation of a two-dimensional electron gas at the surface of annealed SrTiO$_3$ single crystals by scanning tunneling spectroscopy. Phys. Rev. B {\bf 86}, 155425 (2012).
\bibitem{WBao} W. Bao et al., Tunable ($\delta$$\pi$, $\delta$$\pi$)-type antiferromagnetic order in $\alpha$-Fe(Te,Se) superconductors. Phys. Rev. Lett. {\bf 102}, 247001 (2009).
\bibitem{TKChen} T. K. Chen et al., Fe-vacancy order and superconductivity in tetragonal $\beta$-Fe$_{1-x}$Se. Proc. Natl. Acad. Sci. USA {\bf 111}, 63 (2014).

\bibitem{FeTe} T. J. Liu et al.,  From ($\pi$,0) magnetic order to superconductivity with ($\pi$,$\pi$) magnetic resonance in Fe$_{1.02}$Te$_{1-x}$Se$_x$.  Nat.  Mater. {\bf 9}, 716 (2010).


\bibitem{EisakiPRB} H. Eisaki et al., Effect of chemical inhomogeneity in bismuth-based copper oxide superconductors. Phys. Rev. B {\bf 69}, 064512 (2004).


\end {thebibliography}
\end{document}